\documentclass[twocolumn]{aastex63}

\received{September 17, 2020}
\revised{October 5, 2020}
\accepted{November 17, 2020} 
\submitjournal{AJ}

\usepackage[T1]{fontenc}
\usepackage{savesym}
\savesymbol{tablenum}
\usepackage[normalem]{ulem}
\usepackage{graphicx}
\usepackage{natbib}
\bibpunct{(}{)}{;}{a}{}{,}
\usepackage{scrextend}		
\usepackage{multirow}
\usepackage{amssymb} 
\usepackage{gensymb}
\usepackage{xfrac} 
\usepackage{dsfont}
\usepackage{mwe,tikz}
\usepackage[percent]{overpic}
\usepackage{lipsum}
\usepackage{siunitx}
\usepackage{rotating}
\usepackage{xspace}
\usepackage{savesym}
\usepackage{amsmath}
\savesymbol{iint}
\usepackage{txfonts}
\restoresymbol{TXF}{iint}

\newcommand{\ie}{i.e.,\xspace}

\newcommand{\dR}[1]{\frac{\partial #1}{\partial R}}

\newcommand{\sbp}{solid-body precession\xspace}

\newcommand{\q}[1]{\textbf{\textcolor{blue}{#1}}}

\newcommand{\In}{\mathrm{h}}
\newcommand{\Out}{\mathrm{c}}

\shorttitle{Lense-Thirring and supersonic accretion}
\shortauthors{Marcel et al.}

\graphicspath{{./}{figures/}}

\begin{document}

\title{Can Lense-Thirring precession produce QPOs in supersonic accretion flows?}

\correspondingauthor{Grégoire Marcel}
\email{gregoire.marcel@villanova.edu, gregoiremarcel26@gmail.com}

\author[0000-0003-1780-5641]{G. Marcel}
\affiliation{Villanova University, Department of Physics, Villanova, PA 19085, USA}

\author[0000-0002-8247-786X]{J. Neilsen}
\affiliation{Villanova University, Department of Physics, Villanova, PA 19085, USA}

\begin{abstract}

The timing properties of X-ray binaries are still not understood, particularly the presence of quasi-periodic oscillations (QPOs) in their X-ray power spectra. The solid-body regime of Lense-Thirring precession is one prominent model invoked to explain the most common type of QPOs, Type C. However, \sbp requires a specific structure that has not been examined in light of constrained properties of accretion flows. We assume in this paper, as \sbp requires, a disk separated into two flows at a transition radius $r_t$: a cold outer flow and a hot inner flow (playing the role of the corona). We explore the physical structure of both flows using model-independent estimates of accretion parameters. We show that, in order to reproduce the observed X-ray spectra during luminous hard states, the hot flow must accrete at sonic to supersonic speeds, unreachable with typical viscous torques. As a result of this extreme accretion speed (or high $\alpha$ parameter), no region of the disk during these states lies in the `wave-like' regime required for solid-body precession. Furthermore, we expect the flow to align with the black hole spin axis via the Bardeen-Petterson effect inside a radius $r_{\rm break}>r_t$. As a consequence, the hot inner flow cannot exhibit solid body precession --- as currently pictured in the literature --- during luminous hard states. Since Type C QPOs are prevalent in these states, we conclude that this mechanism is unlikely to be responsible for producing Type C QPOs around stellar mass black holes. 

\end{abstract}


\section{Introduction} \label{sec:intro}

X-ray binaries are known to be composed of a companion star and a compact object, a black hole here. These systems are often studied in X-ray, where they exhibit extraordinary changes over timescales varying from seconds to months \citep{Remillard06}. While no precise explanation has been found for their behavior, these long variations in X-rays are usually associated with changes in the accretion flow around the black hole \citep{Yuan14}. The accretion flow is believed to be separated at a transition radius $r_t$ between two flows with different physical properties: a hot, optically thin inner flow, and a cold, optically thick outer flow \citep{Done07}. The physical properties and the thermal structure of the cold flow are fairly understood, and a \citeauthor{SS73} disk (SSD) is the most accurate description \citep{Done07}. However, the nature of the hot flow remains subject to debate: dominant accretion process, magnetic field strength, role of jets and winds, etc. \citep{Yuan14,Marcel18a}.

Changes in the X-ray spectra are also observed on dynamical timescales, over a few seconds or shorter. These timing properties are often studied via X-ray power density spectra \citep{1989ARA&A..27..517V}. Of particular interest are narrow peaks in the power spectra, called quasi-periodic oscillations \citep[QPOs,][]{1977SSRv...20..687M, Zhang13}. QPOs are found over six orders of magnitude in frequency, from mHz to kHz, and are associated with different X-ray spectral states \citep{Motta16}. `Type C' are the most studied type of QPO in the literature; they are commonly detected during the hardest X-ray states, and their peaks vary in frequency in the $0.1-10\,$Hz range \citep{2005ApJ...629..403C}. While there is no consensus explanation for QPOs, their presence and properties tie them to the accretion flow itself, and particularly the hot flow \citep{McClintock06, 2020A&A...640A..18M}.

Different mechanisms have been proposed for the production of type C QPOs, see \citet{2020arXiv200108758I} for a review. We focus here on Lense-Thirring precession \citep[LT,][]{LenseThirring}, which arises in general relativity due to a difference in the orbital frequencies $\Omega_\phi$ and $\Omega_z$. Under certain conditions, this may cause the entire flow to precess as a unique body: the \sbp regime \citep{Fragile07, Ingram09}. In their seminal paper, \citet{Ingram09} showed that the precession frequency of such a flow can match that of type C QPOs, explaining flux fluctuations through a geometrical effect. Solid-body precession is now considered as one of the most promising explanations for the origin of QPOs \citep[][]{2020arXiv200108758I}, but its consistency with the dynamics and the thermal structure of the flow has never been fully addressed. In particular, the model must be consistent with the physical structure of the accretion flow at all times, \ie for any particular state where Type C QPOs are observed.


In this paper we follow up on the origin of QPOs with an analysis of the disk structure and its implications for LT precession. In section \ref{sec:LTprec}, we provide a theoretical background for LT precession. We then describe in section \ref{sec:DiskStructure}, in a model-independent manner, the expected physical structure of the hot flow during luminous hard states. Using these estimates, we address the precession of the disk and discuss the plausibility of solid-body precession during these states in section \ref{sec:radialdistrib}. We discuss and conclude in section~\ref{sec:conclusion}.

\section{Lense-Thirring precession} \label{sec:LTprec}


We envision an X-ray binary with initial disk angular momentum $\vec{L} (R) = L(R) \, \hat{I}$ at radius $R$, $\hat{I}$ defining the normal of the orbital plane. Naturally, $L (R) = \Sigma R^2 \Omega_\phi$, where $\Omega_\phi$ is the azimuthal rotation frequency, and $\Sigma = \rho H$ is the vertical column density, with $H$ the vertical scale height and $\rho$ the midplane density \citep{Frank02}. We use the convention $\Sigma = \rho H$ to be consistent with the work from \citet{Frank02} and \citet{2012ApJ...757L..24N}.

We assume that the disk is misaligned with the black hole spin axis $\hat{k}$, defining an angle $\theta = \text{cos}^{-1} \left( \hat{k} \cdot \hat{I} \right)$. We also define the turbulent viscosity $\nu = \alpha c_s H$, where $\alpha$ is the viscosity parameter and $c_s = \Omega_{\phi} H$ is the sound speed \citep{SS73}. We imagine a quasi-Keplerian flow $\Omega_\phi \approx \sqrt{GM R^{-3}}$, with $G$ the gravitational constant and $M$ the black hole mass. We rescale $r = R / R_g$ and $h = H / R_g$ with respect to the gravitational radius $R_g = GM/c^2$, with $c$ the speed of light in vacuum, and define $\epsilon = h/r$ the aspect ratio of the disk. We label the black hole spin $a>0$, and the LT precession frequency $\vec{\Omega}_p = | \vec{\Omega}_p | \, \hat{k}$ \citep{2012ApJ...757L..24N}.

\subsection{Two torques} \label{sec:twotorques}

\citet{2012ApJ...757L..24N} assumed that the accretion flow was subject to only two torques: the viscous $( \vec{G}_\nu )$ and LT $( \vec{G}_{LT} )$ torques, with
\begin{align}
    G_\nu &= |\vec{G}_{\nu}| = - 2 \pi R \nu \Sigma R^2 \, \dR{\Omega_\phi} , \nonumber \\
    G_{LT} &= |\vec{G}_{LT}| = 2 \pi R H \, | \vec{\Omega}_p \times \vec{L} |  . \nonumber 
\end{align}
%
At a given radius $(r \gg 1 \geq a)$, in a steady state accretion flow, we thus have
\begin{align}
    \frac{G_{LT}}{G_{\nu}} 
    & \simeq \frac{4}{3} \, \frac{a | \mathrm{sin} (\theta ) |}{\alpha \epsilon} \, r^{-3/2} . \label{eq:GnuvsGLT}
\end{align}
When the viscous torque dominates, the accretion flow is not altered by LT precession: the flow remains aligned with its initial angular momentum $\hat{I}$. When the LT torque dominates, however, the flow can precess in different ways.

\subsection{Two regimes} \label{sec:tworegimes}

When the LT torque dominates, \citet{1983MNRAS.202.1181P} showed that there are two possible\footnote{For a more recent and refined study of these regimes see \citet{2011MNRAS.415.2122Z} or \citet{2019ApJ...875....5M}} for example. precession regimes, depending on the damping length of the bending waves in the flow $L_{\mathrm{damp}} = l_{\mathrm{damp}} \, R_g$ \citep[see also][]{1992MNRAS.258..811P, 1999MNRAS.304..557O}, with
\begin{align}
l_{\mathrm{damp}} = \frac{h}{\alpha} = \frac{\epsilon}{\alpha} \, r . \label{eq:ldamp}
\end{align}

When $l_{\mathrm{damp}} > r$, the flow enters a `wave-like regime' where bending waves can propagate radially \citep{1995ApJ...438..841P, 2002MNRAS.337..706L}. When an extended region of the accretion flow ($r_i<r<r_o$) lies in the wave-like regime, it precesses as a unique element, or `solid-body', with a frequency depending primarily on $r_o$ \citep{Fragile07, Ingram09}. The \sbp frequency is significantly lower than the point particle LT precession frequency and matches the observed frequency range of QPOs \citep[][]{2020arXiv200108758I}.

When $l_{\mathrm{damp}} < r$, however, the flow is in the `diffusive regime': the bending waves cannot propagate and all annuli precess locally and independently \citep{2012ApJ...757L..24N}. In this case, the flow will align with the black hole spin wherever $G_{LT}$ is bigger than $G_{\nu}$. Since $G_{LT} / G_{\nu} \propto r^{-3/2}$, LT precession can only occur in the inner regions of the accretion flow and there must be a radius $r_{\mathrm{break}}$ where the viscous torque overtakes the LT torque. The flow is expected to align with the black hole spin inside $r_{\mathrm{break}}$ and with the orbital plane $\hat{I}$ outside of it: the Bardeen-Petterson configuration \citep{1975ApJ...195L..65B}. 

\section{The need for (super)sonic accretion} \label{sec:DiskStructure}

As discussed above, both the torque ratio and the precession regime depend on the viscosity parameter $\alpha$ and the disk aspect ratio $\epsilon$. Thus, LT precession relies on the structure of the accretion flow.
As is done in the LT precession model \citep{Fragile07}, we assume a disk separated into two flows at some radius $r_t$: an inner hot flow and an outer cold SSD \citep{SS73}. In the cold flow, the expected disk aspect ratio is $\epsilon \in [10^{-3}, \, 10^{-2}]$ and the viscosity is due to the magneto-rotational instability (MRI) with $\alpha \in [0.03, \, 1]$ \citep{BH91, 1995ApJ...440..742H, 2018A&A...620A..49S, 2018Natur.554...69T}. In the hot flow, as argued in Sect.~\ref{sec:intro}, most of the physical properties are still subject to debate, but we address them in a model-independent manner in the following subsections. We refer the interested reader to \citet{Frank02} for a more detailed treatment of the equations used in this section. 

\subsection{Formalism}

We define the accretion rate $\dot{M} = - 2 \pi R u_R \Sigma$, with $u_R$ is the radial velocity. We rescale $\dot{m} = \dot{M} c^2 / L_{Edd}$, with $L_{Edd}$ the Eddington luminosity. We also define the vertical optical depth $\tau = \kappa \Sigma $, with $\kappa$ the mean opacity. In the conditions expected in the hot flow, the mean opacity is given by the Thomson regime $\kappa = \sigma_T / m_p$, where $\sigma_T$ is the Stefan-Boltzmann constant and $m_p$ the proton mass.
From the definitions of $\tau$ and $\dot{M}$, we can write the following dimensionless equation for the sonic Mach number $m_s = -u_R / c_s$,
\begin{equation}
    m_s = \frac{\dot{m}}{\epsilon \tau \sqrt{r}} = \dot{m} \, \epsilon^{-1} \tau^{-1} r^{-1/2} . \label{eq:ms} 
\end{equation}
\noindent This allows us to obtain estimates of $m_s$ at any radius $r$, provided values for $\dot{m}$, $\tau$, and $\epsilon$. These equations are not specific to any particular model in the literature \citep{Frank02, Yuan14}, although they would become invalid for a thick disk with $\epsilon \approx 1$.


\subsection{Luminous hard states} \label{sec:HLHS}

From now on, we solely focus on luminous hard states with luminosity $L > 10\% \, L_{Edd}$. This is for two reasons: First, we wish to address the production of type C QPOs, which are preferentially detected in these states \citep{2020A&A...640A..18M}; Second, and more importantly, a high-energy cut-off is crucial to estimate $\tau$, and these cut-offs are only detectable at high luminosity \citep{Motta09}. The vertical optical depth $\tau$ is derived through modeling of the continuum emission with Compton scattering. In luminous hard states, the X-ray continuum is typically fitted with a power-law of slope $\Gamma \simeq 1.6-1.8$ and cut-off $E_c \simeq 50-200\,$keV. Compton models of this continuum give typical values $\tau \sim 1-3$ \citep{1997MNRAS.288..958G, 1998MNRAS.301..435Z, 1999ASPC..161..295B, Ibragimov05, 2006A&A...447..245W, 2006A&A...446..591C, Ingram09}. In what follows, we will use $\tau = 1$. 
Additionally, reaching a luminosity $L = 0.1 \, L_{Edd}$ requires $\dot{M} c^2 \geq L_{Edd}$ for even the most efficient accretion flows \citep[][Figure~2]{Yuan14}.
We therefore require $\dot{m} \geq 1$, and will use $\dot{m} = 1$, the lowest expected value.

In turn, the aspect ratio of the disk $\epsilon$ is usually estimated by solving for the thermal structure of the accretion flow. While it depends on the model and assumptions, it always lies in the range $\epsilon \in [10^{-3}, \, 1]$, with expected values $\epsilon \lesssim 0.01$ in a cold (radiatively efficient) flow, and $\epsilon \gtrsim 0.1$ in a hot (radiatively inefficient) flow. The hot flow is expected to be radiatively-inefficient, and we adopt a standard $\epsilon_{\,\In} = 0.2$ \citep{Ingram09,Marcel18a}.
Typical values for the transition radius in this state lie in the range $r_t = 2-30$, though it depends on the model, method, and assumptions: see for example \citet{2019MNRAS.485.3845D} or \citet{2020ApJ...899...44W} for reflection models and \citet{Marcel19} for a continuum model. We choose $r_t = 10$ (or $R_t = 10\,R_g$) as a central value, so that the hot flow spans $r < r_t = 10$. We note that, according to the \sbp model \citep{Ingram09}, $r_t = 10$ and $\epsilon_{\, \In} = 0.2$ correspond to the $\approx 5\,$Hz type C QPO observed in the luminous hard states considered.

When we combine all these estimates, we obtain the following typical sonic Mach number in the hot flow
\begin{align}
    m_s = 1.6 ~ \Bigg(\frac{0.2}{\epsilon_{\,\In}}\Bigg) \Bigg(\frac{\dot{m}}{1} \Bigg) \Bigg(\frac{1}{\tau} \Bigg) \Bigg(\frac{10}{r}\Bigg)^{1/2} . \label{eq:supersonic}
\end{align}
Thus, for plausible parameters, the accretion speed in the hot flow is expected to reach supersonic values $|u_R| \gtrsim c_s$. We note that $\tau$ is chosen at the lowest expected values, and it is thus possible that the accretion speed be as low as $m_s \simeq 0.5$, \ie the flow is not supersonic.
Moreover, the accretion timescale $t_{acc} = R/|u_R|$ is expected to be
\begin{align}
    t_{acc} = 0.5 ~ \Bigg(\frac{1}{\dot{m}} \Bigg) \Bigg(\frac{\tau}{1} \Bigg) \Bigg(\frac{r}{10}\Bigg)^{1/2} \, \Bigg(\frac{2 \pi}{\Omega_\phi} \Bigg) 
    \label{eq:tacc} ,
\end{align}
\noindent shorter than the actual orbital period $( 2 \pi / \Omega_\phi )$ throughout the entire hot flow. Such a short accretion timescale undoubtedly has an important impact on the propagation of fluctuations, warps, and instabilities in the accretion flow, but their study is beyond the scope of this work. 

\subsection{Production of (super)sonic accretion} \label{sec:supersonic}

Let us imagine that this accretion is generated only via turbulent viscosity $\nu = \alpha_{\In} c_s H$. We can write $\nu \Sigma \simeq \dot{M} / (3 \pi)$ \citep{Frank02}, leading to $m_s \simeq 1.5 \, \alpha_{\In} \epsilon_{\In}$, and thus
\begin{align}
    \alpha_{\In} = 5.3 ~ \Bigg(\frac{0.2}{\epsilon_{\,\In}}\Bigg)^2 \Bigg(\frac{\dot{m}}{1} \Bigg) \Bigg(\frac{1}{\tau} \Bigg) \Bigg(\frac{10}{r}\Bigg)^{1/2} . \label{eq:bigalpha}
\end{align}
This shows that an extremely high viscosity parameter $\alpha_{\In} > 1$ would be required in the hot flow to reproduce the luminous hard states spectra. The MRI, believed to be at the origin of viscous accretion \citep{BH91}, can produce high viscosities in the presence of a strong large scale magnetic field \citep{1995ApJ...440..742H, Salvesen16}. However, the case $\alpha_{\In} \gg 1$ cannot be produced by MRI turbulence alone; instead, magnetic outflows are required to enhance the effective $\alpha$ parameter through large scale magnetic torques \citep[see for example][]{2018A&A...620A..49S}. We will thus consider two cases.

In the first case, $\alpha$ is produced by MRI turbulence only, reaching a maximum value $\alpha_{\In} = 1$. In this case, the only relevant torques are viscous and LT, and we can estimate the physical structure of the flow using $\alpha_{\In}=1$ and $G_{\nu} (\alpha_{\In})$. 

In the second case, the physical structure is established by three torques: a turbulent torque, a LT torque, and a torque due to the outflow. To explore this scenario, we use the solution labeled jet-emitting disk \citep[JED,][]{Ferreira93a,Ferreira93b}. We assume the presence of a strong vertical magnetic field, driving an outflow that acts on the flow itself through the magnetic torque $\vec{G}_{mag} = \vec{R} \times \left( \vec{J} \times \vec{B} \right)$, where $\vec{J}$ and $\vec{B}$ are the magnetic current and field \citep{BP82, Ferreira97}. The two accretion torques, $G_{\nu}$ and $G_{mag}$, are commonly parameterized by their $\alpha$ prescriptions\footnote{See caveats to this approach in Section 4.1.}, respectively $\alpha_\nu$ and $\alpha_{mag}$. The magnetic field is at equipartition in this solution, leading to $\alpha_\nu \simeq 1$ \citep{Salvesen16}. In turn, $\alpha_{mag}$ can be given by the disk sonic Mach number $m_s = 1.5 (\alpha_{mag} + \alpha_\nu) \epsilon$. When $\alpha_{mag} \gg \alpha_{\nu}$, this gives $\alpha_{mag} \simeq m_s/( 1.5 \epsilon )$ in the JED region, where $m_s$ is a parameter \citep{Marcel18a}. When coupled with a SSD as the cold flow (or SAD for standard accretion disk), the solution is labeled JEDSAD and can reproduce the X-ray spectra of luminous hard states \citep{Marcel18a,Marcel18b}. This solution is thus perfectly suited to compare with our hot and cold flows. 

\section{Physical structure} \label{sec:radialdistrib}

\subsection{Three different cases} \label{sec:3sols}

To address the physical structure of the disk, we explore both analytical and numerical estimates described above.

We first perform analytical estimates in the case of viscous-only accretion. We set all parameters to the common values set forth in previous sections: $\alpha_{\Out} = 0.1$ and $\epsilon_{\Out} = 0.01$ in the cold flow, and $\epsilon_{\,\In} = 0.2$ and $\alpha_{\In} = 1$ in the hot flow (see Sect.~\ref{sec:HLHS} and \ref{sec:supersonic}). In a more realistic model, these values would evolve with radius, but we only wish here to have the simplest estimates. This solution is shown in red in Fig.~\ref{fig:estimates}.

We then use the more realistic solution calculated with the JEDSAD model. We freeze the parameters to the typical values of this letter, \ie the accretion rate at the innermost stable circular orbit $\dot{m}_{in} = 1$, the transition radius $r_J = 10$, the outer disk viscosity $\alpha = 0.1$, and the inner disk Mach number $m_s=1.5$ \citep[for all parameters of the model, see][]{Marcel18a,Marcel18b}. This solution is known to reproduce the luminous hard states considered \citep{Marcel18b}. However, while it is common to reduce all angular momentum transport to the effective viscosity $\alpha$, this is not always valid \citep{2018A&A...620A..49S}. For instance, it is unclear how the torque from the outflow would affect the bending waves or even the LT torque itself. For that reason, we consider two cases in the JEDSAD solution.
Our first numerical solution considers the total torque $G_{tot} = G_{mag} + G_{\nu}$, assuming it counteracts the LT torque $G_{LT}$, and that the bending waves are damped through the total $\alpha$-prescription $\alpha_{tot} = \alpha_{mag} + \alpha_\nu$. This solution is shown in solid blue in Fig.~\ref{fig:estimates}. Our second numerical solution assumes that only the diffusive torque $G_{\nu}$ counteracts the LT torque and that the waves are damped only through the true viscosity $\alpha_\nu$. This solution is shown in dotted blue in Fig.~\ref{fig:estimates}. We note that the two JEDSAD solutions differ only in the hot flow, where the magnetic torque dominates. As discussed above, it is also possible that the outflow torque could act with the LT torque, \ie enhancing precession, but as we discuss in Section~\ref{sec:conclusion}, little is known about this possibility and the details are beyond the scope of this paper.

We detail in Sect.~\ref{sec:outerflow} and \ref{sec:innerflow} the physical structure in each case, assuming $a=0.9$ and $\theta = 30 \degree$, \ie sin$(\theta) = 0.5$.

\subsection{Outer flow} \label{sec:outerflow}

\begin{figure}[ht!]
\includegraphics[width=1.0\linewidth]{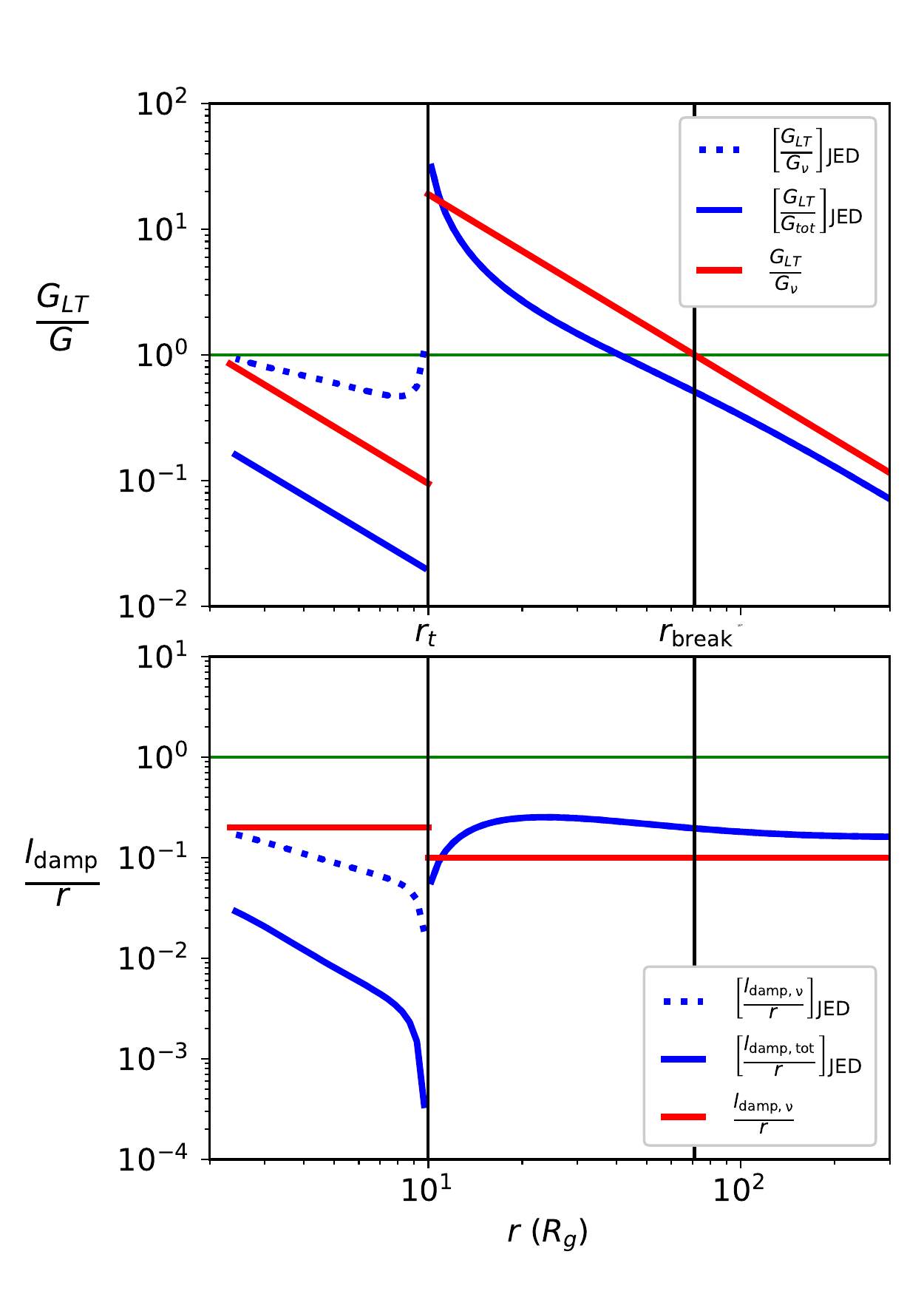}
\caption{Top: radial distribution of $G_{LT} / G$ in the simple accretion flow (red) and the JEDSAD solutions (blue). Bottom: radial distribution of $l_{\mathrm{damp}} / r$ in the simple accretion flow (red) and the JEDSAD solutions (blue). The vertical lines show the different radii in the accretion flow; in this example, $r_t = 10$ and $r_{\mathrm{break}} \simeq 71$.}
\label{fig:estimates}
\end{figure}

Let us first focus on the cold flow, \ie outside of $r_t = 10$. Using the estimates performed previously for the analytical solution we have
\begin{align}
    \left[ \frac{G_{LT}}{G_{\nu}} \right]_{\Out} & \simeq 19 ~
    \Bigg(\frac{0.01}{\epsilon_{\Out}}\Bigg) \Bigg(\frac{0.1}{\alpha_{\Out}}\Bigg) \Bigg(\frac{a}{0.9}\Bigg) \Bigg(\frac{| \mathrm{sin}(\theta)| }{0.5}\Bigg) \Bigg(\frac{10}{r}\Bigg)^{3/2} \nonumber , \\
    \left[ \frac{l_{\mathrm{damp}}}{r} \right]_{\Out} & \simeq \frac{1}{10} ~ \Bigg(\frac{\epsilon_{\Out}}{0.01}\Bigg) \Bigg(\frac{0.1}{\alpha_{\Out}}\Bigg) . \nonumber
\end{align} 
This is illustrated on the top panel of Figure~\ref{fig:estimates}, in red. For the typical values considered, the LT torque dominates in the inner region of the cold flow, between $r_t$ and $r_{\mathrm{break}} = 71$. Outside $r_{\mathrm{break}}$, the LT torque is negligible. The entire flow is in the diffusive regime, \ie $l_{\mathrm{damp}} < r$.

This configuration is that of a warped (or torn for large angles) disk: the accretion flow is aligned with the black hole spin axis $(\hat{k})$ inside $r_{\mathrm{break}}$, but it is aligned with its initial orbital plane $(\hat{I})$ outside $r_{\mathrm{break}}$. We estimate a warping (or tearing) radius
\begin{align}
    r_{\mathrm{break}} \simeq 71 ~ \Bigg( \frac{0.01}{\epsilon_{\, \Out }} \Bigg)^{2/3} \Bigg( \frac{0.1}{\alpha_{\Out}} \Bigg)^{2/3} \Bigg( \frac{a}{0.9} \Bigg)^{2/3} \Bigg( \frac{| \mathrm{sin}(\theta)| }{0.5} \Bigg)^{2/3} , \label{eq:Rbreak}
\end{align}
\noindent a value smaller than that of active galactic nuclei \citep{2012ApJ...757L..24N} because of the higher expected value of $\alpha_{\Out}$ in X-ray binaries. For the assumptions made here, the warping radius will be larger than $r_t = 10$ as long as $\theta \gtrsim \theta_0 \simeq 1.5 \degree$. We assumed that $\alpha_{\Out}$ and $\epsilon_{\,\Out}$ are radially constant, but both are expected to decrease with radius, increasing $r_{\mathrm{break}}$ and decreasing the minimum angle $\theta_0$. One thus expects $r_{\mathrm{break}} > r_t$ in the luminous hard states. 

In the JEDSAD solutions (in blue), as in the analytical case, the entire cold flow lies in the diffusive regime, and the LT torque dominates the inner region of the cold flow out to a similar radius $r \simeq 41$. 

\subsection{Inner flow} \label{sec:innerflow}

We now focus on the hot flow, where analytical values give
\begin{align}
    \left[ \frac{G_{LT}}{G_{\nu}} \right]_{\In} & \simeq \frac{1}{11} ~ \Bigg(\frac{0.2}{\epsilon_{\In}}\Bigg) \Bigg(\frac{1}{\alpha_{\In}}\Bigg) \Bigg(\frac{a}{0.9}\Bigg) \Bigg(\frac{| \mathrm{sin}(\theta)| }{0.5}\Bigg) \Bigg(\frac{10}{r}\Bigg)^{3/2} \nonumber , \\
    \left[ \frac{l_{\mathrm{damp}}}{r} \right]_{\In} & \simeq \frac{1}{5} ~ \Bigg(\frac{\epsilon_{\In}}{0.2}\Bigg) \Bigg(\frac{1}{\alpha_{\In}}\Bigg) . \nonumber
\end{align}
\noindent We show the radial distributions in Figure~\ref{fig:estimates}, where the LT torque is smaller than the viscous torque. The damping length of bending waves $l_{\mathrm{damp}}$ is also always smaller than the radius $r$. The disk remains aligned with its initial angular momentum, \ie the one it inherits at $r_t$: the spin axis of the black hole $\hat{k}$. We note that we have made these calculations with $\alpha_{\In} = 1$, when the expected value was in fact $\alpha_{\In} \simeq 5$. A higher value of $\alpha_{\In}$ would increase the viscous torque and decrease the damping length of the bending waves: the inner flow would still be dominated by the viscous torque and be diffusive, as seen for the JEDSAD solution (below).

In the numerical solution, the JEDSAD, there are now two cases: when the total torque is relevant and when only the viscous torque is relevant. When we consider the total torque (solid blue) the LT torque is completely negligible $G_{LT} \ll G_{tot}$. Moreover, the damping length of the bending waves is much smaller due to the high value $\alpha_{tot} > 1$. When we only consider the viscous torque (dotted blue) we retrieve a solution similar to the analytical $\alpha_{\In} = 1$ case: the LT torque is negligible and the disk is diffusive. The major difference resides at the transition between the two accretion flows, but this is natural because this transition requires a sonic point at $r_J = r_t$ in the JEDSAD model, best described by $\epsilon (r_J) = 0$ \citep{Marcel18b}.

In all three cases, one analytical and two numerical, the LT torque is negligible in the inner flow and bending waves have a damping length too short to support the wave-like regime.

\subsection{Disk geometry} \label{sec:geometry}

By the above arguments, we expect the disk in the luminous hard states to be composed of three zones separated by $r_t$ and $r_{\mathrm{break}}$ (Sect.~\ref{sec:LTprec}).
We illustrate in Figure~\ref{fig:cas2} this geometrical configuration. The hot flow extends out to $r_t$ (in yellow), while the cold flow is separated into two parts by $r_{\mathrm{break}}$ (in green and violet). The outermost part is aligned with the orbital plane, while the Bardeen-Petterson effect causes the inner portion of the cold flow to align with the black hole spin axis $(\hat{k})$. Interestingly, the hot flow is aligned with the black hole spin axis even if the LT torque is negligible $G_{LT} \ll G_{\nu}$.

\subsection{Solid body precession} \label{sec:sbp}

\begin{figure}[ht!]
\includegraphics[width=1.0\linewidth]{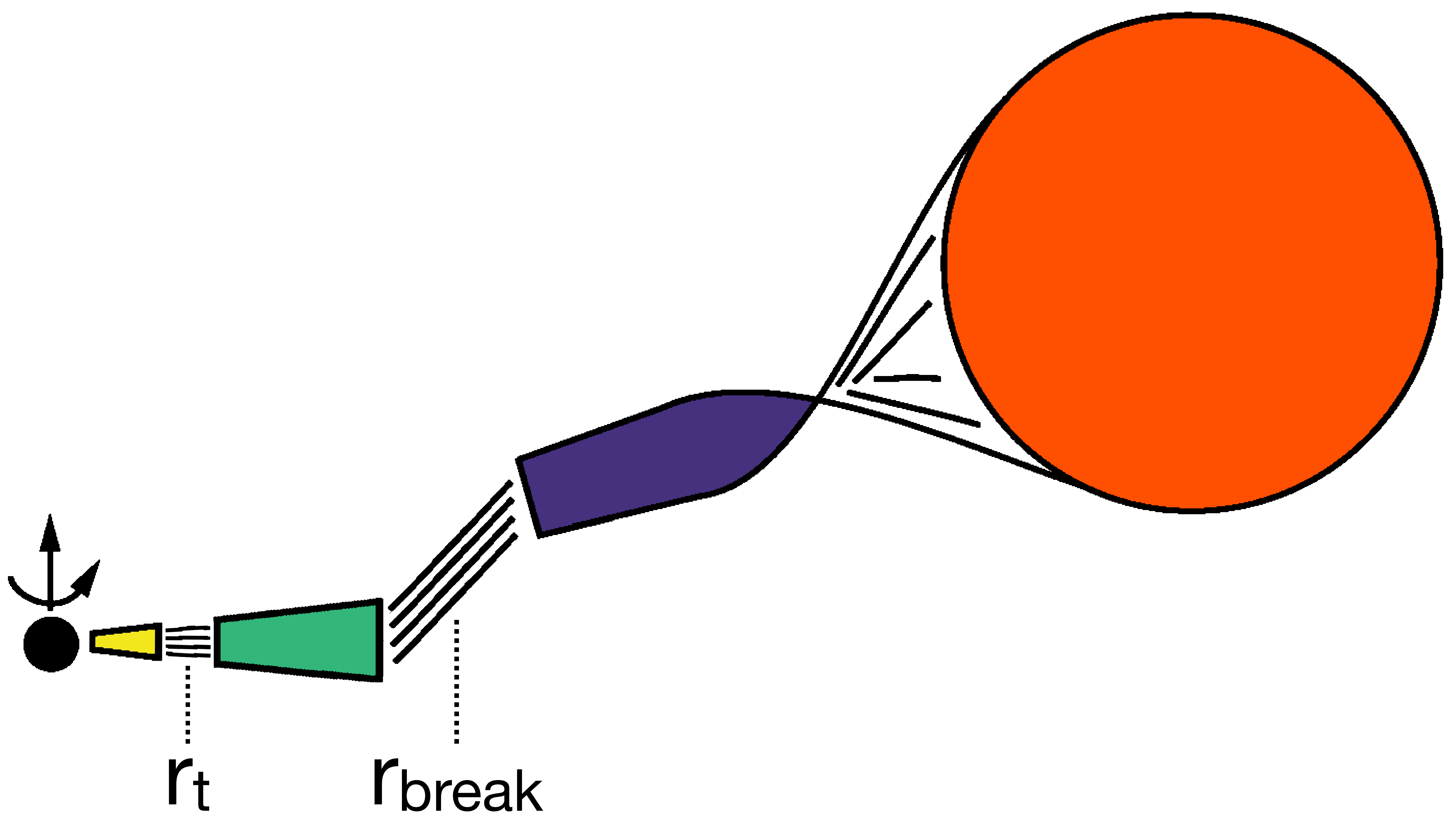}
\caption{Expected schematic configuration of the accretion flow in the luminous hard states. From left to right: the black hole (black), the hot flow (yellow) until $r_t$, the cold flow where it is subject to LT-precession (green) until $r_{\mathrm{break}}$, the cold flow where it is not subject to precession (violet), and the companion star (orange). This Figure has been adapted from \citep{2001ApJ...553..955F}.}
\label{fig:cas2}
\end{figure}

\citet{Ingram09} suggested that \sbp of the hot flow could produce type C QPOs. We know that this regime requires the hot flow to be dominated by $G_{LT}$ and be in the wave-like regime (see Sect.~\ref{sec:tworegimes}). However, we clearly establish that the LT torque is negligible and the flow is diffusive in the key state that is the high-luminosity hard state (see Fig.~\ref{fig:estimates}). It is hard to imagine a reasonable combination of parameters that would enable the accretion flow to meet both conditions while remaining consistent with observations in this state (see Sect.~\ref{sec:HLHS}). Moreover, the cold flow is in the diffusive regime and, similarly, cannot undergo any \sbp . Therefore, the inner region of the cold flow aligns with the black hole spin axis $\hat{I} = \hat{k}$ (see Figure~\ref{fig:cas2}), forcing $\theta = 0\degree$, and thus $G_{LT}=0$ in the hot flow. 

As a result, the \sbp regime described in \citet{Ingram09} cannot develop in the hot flow during luminous hard states. Since a significant fraction of type C QPOs are observed in such luminous hard states \citep{2020A&A...640A..18M}, the \sbp model is unlikely to be at the origin of type C QPOs.

\section{Discussion and conclusion} \label{sec:conclusion}

In light of the accretion flow physical structure, we investigate whether or not the \sbp model is a viable mechanism for Type C QPOs. We focus on luminous hard states ($L > 10\% \, L_{Edd}$), and assume a disk separated into two different flows at a transition radius $r_t$: a hot and a cold flow, as was done by \citet{Ingram09}.

We show that a high accretion speed $|u_R| \gtrsim c_s$ is required in the hot flow during the luminous hard states. This speed would require an unexpectedly high viscosity $\alpha \gtrsim 5$ in the hot flow, a value unreachable with viscous torques alone.

We then study three different cases: one analytical and two numerical with the JEDSAD solution, and show that the entire disk is expected to be in the diffusive regime in each case. Moreover, the LT torque is negligible in the hot flow compared to the viscous torque in all three cases (regardless of initial alignment), and it will not exhibit \sbp regime described by \citet{Ingram09}. Since most type C QPOs are observed during the luminous hard states considered in this letter, we conclude that QPOs cannot originate from the solid-body precession of the hot flow. This mechanism --- in its current form in the literature --- is therefore not responsible for the Type C QPOs observed in high-luminosity hard states. Since a significant fraction of Type C QPOs are found in these states \citep{2020A&A...640A..18M}, we conclude that Type C QPOs cannot originate in \sbp of the hot flow.
We also show that we expect the outer cold flow to be warped (or torn for large angles) into two different planes at a radius $r_{\mathrm{break}} > r_t$: the outer parts remain aligned with the initial angular momentum, while the inner parts align with the black hole spin axis \citep{1975ApJ...195L..65B, 2012ApJ...757L..24N, 2020MNRAS.tmp..707L}.

However, the exact configuration and its implications for LT precession remain to be studied. The broader picture presented in this paper, in which high accretion speeds and large effective viscosity parameters are required to explain the observational characteristics of luminous hard states, has not been integrated into precession models. Of particular interest is the jet: while we have shown that the LT torque is negligible, additional magnetic torques on the disk (e.g., from jets emitted by the hot inner flow) could potentially drive precession and produce unfamiliar configurations of the flow. Furthermore, we assume that the viscosity $\alpha$ is isotropic, but recent work suggest a more complicated picture (\citealt{2013ApJ...777...21S},  \citealt{2014ApJ...796..103M}, but see \citealt{2015MNRAS.450.2459N}). 

In conclusion, while there is evidence that type C QPOs are caused by a geometric effect \citep{2020arXiv200108758I}, we have shown that the \sbp caused by the LT mechanism cannot, in its current form, be the key to produce the actual precession. However, we expect that the extreme accretion speed, materializing in accretion timescale shorter than the actual orbital period (see Sect.~\ref{sec:supersonic}), will prevent the plasma from counteracting any misalignment or precession. In other words, the hot flow could precess as a single flow, but the LT mechanism is unlikely to contribute significantly to the precession. Besides, given the success of accretion-ejection models to reproduce the multi-wavelength behavior of X-ray binaries \citep{Marcel19}, we believe that the answers to the production of QPOs could lie in the powerful (possibly disk-driven) ejections observed in the system.

\section*{\textsc{acknowledgements}}
 We would like to thank the anonymous referee for a careful reading of the paper. GM would like to thank P.-O. Petrucci, C. Fragile, and N. Scepi for important advice that improved this paper, as well as A. Ingram, and J. Ferreira for very fruitful discussions.

\bibliography{Research.bib}
\bibliographystyle{aasjournal}

\end{document}